\documentclass[twocolumn]{aastex7}

\newcommand\multimoon{\texttt{MultiMoon}}
\newcommand\jt{$J_2$}
\newcommand\ct{$C_{22}$}
\newcommand{\trackchange}[1]{\textcolor{black}{#1}}

\usepackage[utf8]{inputenc}
\usepackage{tipa}
\usepackage{combelow}
\usepackage{savesym}
\savesymbol{tablenum}
\usepackage{siunitx}
\restoresymbol{SIX}{tablenum}
\usepackage{breqn}
\usepackage{tabularx}
\usepackage{amsmath}
\usepackage{paralist}

\shorttitle{Beyond Point Masses. V. Quaoar-Weywot}
\shortauthors{Proudfoot et al.}

\begin{document}

\title{Beyond Point Masses. V. Weywot's Non-Keplerian Orbit}

\author[orcid=0000-0002-1788-870X,sname='Proudfoot']{Benjamin Proudfoot}
\affiliation{Florida Space Institute, University of Central Florida, 12354 Research Parkway, Orlando, FL 32826, USA}
\email[show]{benp175@gmail.com}

\author[orcid=0000-0002-8296-6540, sname='Grundy']{Will Grundy} 
\affiliation{Lowell Observatory, 1400 W Mars Hill Rd, Flagstaff, AZ 86001, USA}
\affiliation{Northern Arizona University, Department of Astronomy \& Planetary Science, PO Box 6010, Flagstaff, AZ 86011, USA}
\email{grundy@lowell.edu}

\author[orcid=0000-0003-1080-9770, sname='Ragozzine']{Darin Ragozzine} 
\affiliation{Brigham Young University Department of Physics \& Astronomy, N283 ESC, Brigham Young University, Provo, UT 84602, USA}
\email{darin_ragozzine@byu.edu}

\author[orcid=0000-0003-2132-7769, sname='Fern\'{a}ndez-Valenzuela']{Estela Fern\'{a}ndez-Valenzuela}
\affiliation{Florida Space Institute, University of Central Florida, 12354 Research Parkway, Orlando, FL 32826, USA}
\email{estela@ucf.edu}

\begin{abstract}
We present a detailed dynamical analysis of the Quaoar–Weywot system based on nearly 20 years of high-precision astrometric data, including new HST observations and stellar occultations. Our study reveals that Weywot's orbit deviates significantly from a purely Keplerian model, requiring the inclusion of Quaoar’s non-spherical gravitational field and center-of-body--center-of-light (COB-COL) offsets in our orbit models. We place a robust upper limit on Weywot’s orbital eccentricity ($e<0.02$), substantially lower than previous estimates, which has important implications for the strength of mean motion resonances (MMRs) acting on Quaoar’s ring system. \trackchange{Under the assumption that Quaoar's rings lie in its equatorial plane, we} detect Quaoar's dynamical oblateness, \jt, at $\sim$2$\sigma$ confidence. \trackchange{The low \jt{} value found under that assumption implies} Quaoar is differentiated, with a total bulk density of $1751\pm13$ (stat.) kg m$^{-3}$. Additionally, we detect significant COB-COL offsets likely arising from latitudinal albedo variations across Quaoar’s surface. These offsets are necessary to achieve a statistically robust orbit fit and highlight the importance of accounting for surface heterogeneity when modeling the orbits of dwarf planet moons. These findings improve our understanding of Quaoar’s interior and surface while providing key insights into the stability and confinement mechanisms of its rings.

\end{abstract}

\keywords{\uat{Trans-Neptunian objects}{1705}, \uat{Asteroid satellites}{2207}, \uat{Natural satellite dynamics}{2212}, \uat{Dwarf planets}{419}, \uat{Planetary interior}{1248}, \uat{Orbit determination}{1175}}

\begin{figure*}
    \centering\includegraphics[width=0.99\textwidth]{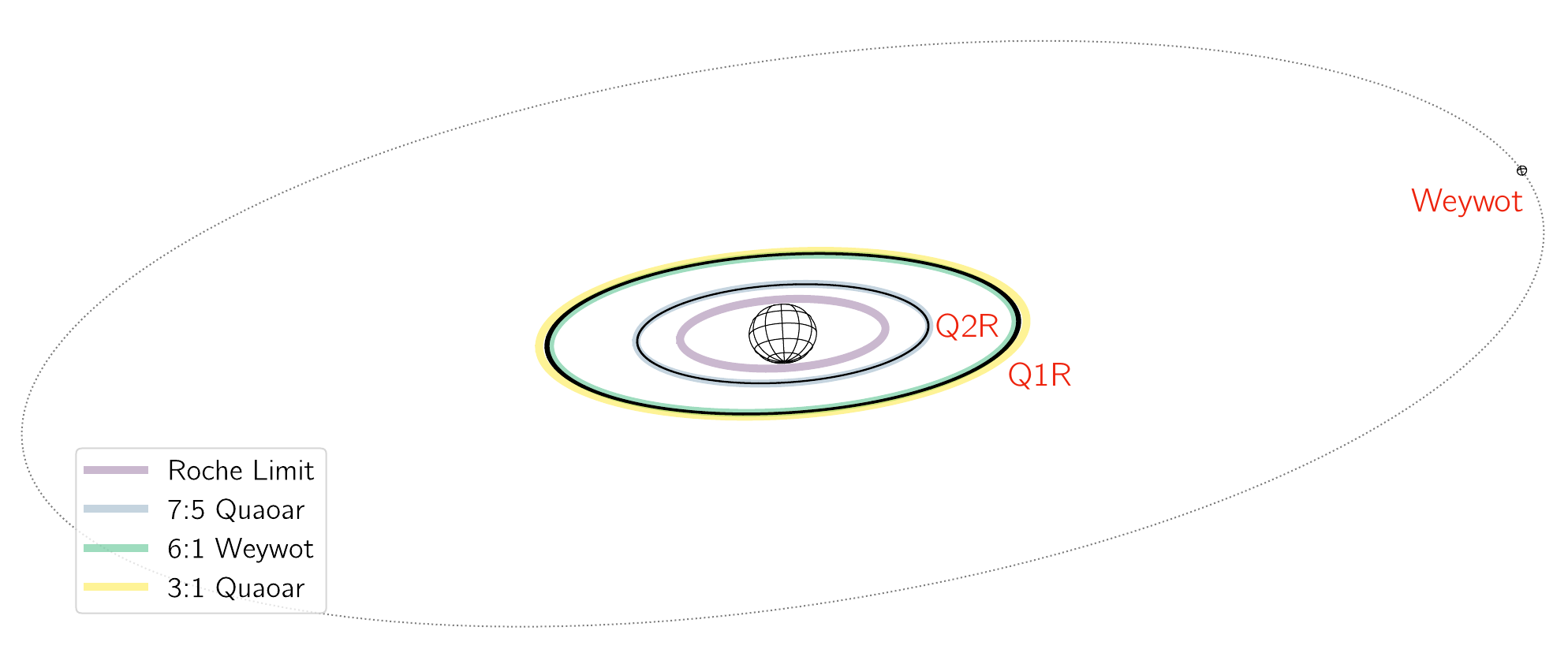}
    \caption{The architecture of the Quaoar system, as seen from Earth, shown to scale on 2006 Sep 21 12:00 UT. Locations of various spin-orbit and mean motion resonances are shown as colored ellipses. Quaoar pole orientation, ring width, and ring orientation are based on \citet{proudfoot2025jwst}. Quaoar's shape is based on \citet{margoti2024quaoarshape}, although the rotational phase is unknown, but selected to be at light curve maximum. Weywot's diameter from Fernandez-Valenzuela et al. (2025, in prep). Weywot orbit from this work. }
    \label{fig:schematic}
\end{figure*}

\section{Introduction}
\label{sec:intro}
The dwarf planet Quaoar is one of the most intriguing bodies yet discovered in the trans-Neptunian region. Like other large trans-Neptunian objects (TNOs), Quaoar hosts a relatively small moon---Weywot---discovered almost two decades ago in data from the Hubble Space Telescope (HST) \citep{brown2007moondiscovery}. In addition, recent observations of stellar occultations by Quaoar show the existence of two rings orbiting outside of Quaoar's Roche limit \citep{morgado2023dense,pereira2023two}. \trackchange{Repeated observations of these enigmatic rings have revealed that both appear to have substantial azimuthal structure, which have been interpreted as arc-like structures embedded within the rings. This is similar to the ring arcs seen in Neptune's Adams ring. These arcs are} possibly caused by resonances with either Quaoar's spin orbit resonances (SORs) or Weywot's mean motion resonances (MMRs). Understanding how these rings formed, are confined, and \trackchange{have not} accreted/dissipated is the subject of intense interest \citep[e.g.][]{rodriguez2023dynamical,ikeya2024gravitational}. Given this interest, characterization of Quaoar's dynamical environment is crucial. 

Quaoar's dynamical environment is dominated by both Weywot's orbital influence (especially near MMRs) and Quaoar's non-spherical gravitational field (especially near SORs). The ring system appears to be embedded \trackchange{within/near several of these resonances}, with the outer ring---referred to as Q1R---located near Weywot's 6:1 MMR and Quaoar's 3:1 SOR \citep{morgado2023dense}. Likewise, the inner ring---Q2R---is located near Weywot's 12:1 MMR and Quaoar's 7:5 SOR \citep[][]{pereira2023two,proudfoot2025jwst}. We show a to scale schematic of the Quaoar system, \trackchange{along with its SORs and MMRs}, in Figure \ref{fig:schematic}. \trackchange{For a higher resolution look at the resonances near Q1R, we refer the reader to Figure 7.} Understanding the relative strength and location of these resonances is central to understanding the ring system. 

Weywot's MMRs are dependent on its mass and orbital parameters, while Quaoar's SORs are dependent on the shape of Quaoar's presumably triaxial gravitational field. Unfortunately, to date, strong constraints have not been able to be placed on almost all relevant parameters. Importantly, Weywot's orbit has yet to be fully constrained. Since the first publications on Weywot's orbit, \trackchange{almost every subsequent study has yielded notably different system mass, orbital period, and/or eccentricity estimates \citep[e.g.][]{fraser2010quaoar,vachier2012determination,morgado2023dense}, underscoring the difficulty of achieving a consistent solution. These values are }strongly related to the locations and strengths of MMRs. Likewise, the shape of Quaoar's gravitational field remains unconstrained. Some three-dimensional shape models of Quaoar have been published \citep{margoti2024quaoarshape,kiss2024visible}, but these only place bounds on the shape of Quaoar's gravitational field. 

Characterizing the dynamical environment of Quaoar is best achieved by precise astrometric tracking of Weywot's position. This allows Weywot to act as a probe of Quaoar's gravitational field, while precisely characterizing the dynamical contributions from Weywot. This allows us to simultaneously constrain the properties of both Quaoar's SORs and Weywot's MMRs. 

In this work, we parameterize Quaoar's nonspherical gravitational field using a spherical harmonic expansion. The nonspherical potential ($U$) of Quaoar (with mass $M$) at a distance ($r$) can be written to quadrupole-order as:

\begin{dmath}\label{eqn:potential}
    U(r,\theta,\phi) = -\frac{GM}{r} \left[
    1 - J_2 \left( \frac{R}{r} \right) ^2 \left( \frac{3}{2}\sin^2\theta - \frac{1}{2} \right) 
    + C_{22} \left( \frac{R}{r} \right) ^2\cos^2\theta\sin2\phi + 
    \mathcal{O} \left( r^{-3} \right) \right]
\end{dmath}

\noindent where \jt{} and \ct{} are the quadrupole-order gravitational harmonic coefficients, $\theta$ is the body-fixed latitude, $\phi$ is the body-fixed longitude, and $R$ is the ``reference'' radius \citep[][]{1995geph.conf....1Y, scheeres2000evaluation}\footnote{Throughout this paper, to simplify discussion and to provide comparison to the literature, we assume Quaoar's reference radius is 549 km, the volume equivalent radius found in a recent study \citep{margoti2024quaoarshape}.}.
The term \jt{} corresponds to the oblate potential of Quaoar, which causes ``splitting'' of Weywot's MMRs into distinct subresonances. Measuring Quaoar's \jt{} is therefore important for understanding the dynamical state of Quaoar's rings.
In contrast with \jt, \ct{} corresponds to a rotating prolate potential aligned with Quaoar's long axis. The value of \ct{} is directly related to the strength of Quaoar's SORs. Both \jt{} and \ct{} are dependent on a body's shape and internal density structure. If a body's exterior figure is well-known, the harmonics place strong constraints on possible internal density structures \citep[e.g.][]{proudfoot2024bpm3}. 


In this work, we present a comprehensive orbital analysis of the Quaoar–Weywot system. With nearly two decades of astrometric data, we use an advanced non-Keplerian orbit fitting scheme to better understand Quaoar's shape, interior, and the dynamical environment of its ring system. In Section \ref{sec:obs}, we present new observations of Quaoar-Weywot and reanalyze archival observations. Then, in Section \ref{sec:models}, we describe the different orbital models we use to fit our data, the results of which are presented in Section \ref{sec:results}. We further describe the implications for Quaoar's shape and interior in Section \ref{sec:interior}, and for the albedo distribution across Quaoar's surface in Section \ref{sec:offsets}. We then examine Quaoar's ring dynamics in detail in Section \ref{sec:rings} to finally conclude in Section \ref{sec:conclusions}.

\section{Observations}
\label{sec:obs}
Based on orbit fits available in the literature and updated online, past works have found significantly different orbital configurations for Weywot, even with similar data sets, presenting significant disagreements on the period and eccentricity \citep{fraser2010quaoar,fraser2013mass,vachier2012determination,morgado2023dense}. Motivated by these contradictory conclusions, we have reanalyzed all past HST observations and combine them with a set of 10 new observations since 2019. 

\begin{deluxetable*}{cccCCCC}
\tablewidth{\textwidth}
\tablecaption{Observed Astrometric Positions of Weywot}
\tablehead{
Julian Date & Date & Telescope/Instrument & \Delta \alpha \cos{\delta} & \Delta \delta & \sigma_{\Delta \alpha \cos{\delta}} & \sigma_{\Delta \delta} \\
 & & & ('') & ('') & ('') & ('')
}
\startdata
2453781.39621 & 2006/02/14 & HST/ACS-HRC & +0.32879 & -0.13111 & 0.00797 & 0.00253 \\
2454179.14096 & 2007/03/19 & HST/WFPC2 &+ 0.31771 & -0.12884 & 0.00188 & 0.00264 \\
2454535.70523 & 2008/03/10 & HST/WFPC2 & -0.39224 & -0.02512 & 0.00396 & 0.00363 \\
2454540.56322 & 2008/03/15 & HST/WFPC2 & +0.40824 & -0.07205 & 0.00371 & 0.00144 \\
2454546.18613 & 2008/03/20 & HST/WFPC2 & -0.36223 & +0.12295 & 0.00332 & 0.00370 \\
2455719.82856 & 2011/06/07 & Keck/NIRC2 & +0.17000 & -0.16000 & 0.01000 & 0.01000 \\
2455719.89737 & 2011/06/07 & Keck/NIRC2 & +0.18000 & -0.16000 & 0.01000 & 0.01000 \\
2455719.95760 & 2011/06/07 & Keck/NIRC2 & +0.19000 & -0.17000 & 0.01000 & 0.01000 \\
2455720.01759 & 2011/06/07 & Keck/NIRC2 & +0.20000 & -0.16000 & 0.01000 & 0.01000 \\
2458700.22561 & 2019/08/04 & Occultation & -0.41809 & -0.06536 & 0.00396 & 0.00446 \\
2459741.88464 & 2022/06/11 & Occultation & -0.21770 & +0.12770 & 0.00350 & 0.00310 \\
2460091.16189 & 2023/05/26 & Occultation & -0.38953 & +0.03468 & 0.00396 & 0.00446 \\
2460108.25987 & 2023/06/12 & HST/WFC3 & +0.13864 & -0.13783 & 0.00233 & 0.00281 \\
2460117.82816 & 2023/06/22 & Occultation & -0.39599 & -0.11332 & 0.00300 & 0.00300 \\
2460452.27196 & 2024/05/21 & HST/WFC3 & -0.42830 & -0.01869 & 0.00136 & 0.00100 \\
2460484.13879 & 2024/06/22 & HST/WFC3 & +0.42755 & +0.08743 & 0.00100 & 0.00100 \\
2460764.49131 & 2025/03/29 & HST/WFC3 & -0.36510 & -0.12864 & 0.00100 & 0.00127 \\
2460770.71983 & 2025/04/05 & HST/WFC3 & +0.36846 & +0.13664 & 0.00173 & 0.00100 \\
\enddata
\tablecomments{$x$ and $y$ correspond to Right Ascension and Declination, respectively. Uncertainties on $\Delta x,y$ have noise floors of 1 mas and 3 mas for telescope and occultation-derived observations, respectively.}
\label{tab:observations}
\end{deluxetable*}

\subsection{Reanalysis of Archival Observations}
\label{sec:reanalysis}
Between 2006 and 2008, five epochs of observations were taken by HST using both the Advanced Camera for Surveys High Resolution Channel (ACS/HRC) and the Wide Field and Planetary Camera 2 (WFPC2). \trackchange{An earlier analysis of this dataset \citep{fraser2010quaoar} reported an eccentric orbit for Weywot. Using the ACS/HRC and WFPC2 pipelines developed to derive relative astrometry from HST images for other binary systems \citep[e.g.,][]{grundy200842355,grundy2009mutual}, we re-extracted the astrometry, obtaining positions that differ by up to 100 mas from the originally published values. Differences of this magnitude can have a substantial effect on the derived orbital solution.}

Although our techniques have significant heritage and have been widely validated on many different TNO binaries \citep[][and references therein]{grundy2019mutual}, a selection of one pipeline over another would be \textit{ad hoc}. Given the abundance of data, however, we can explicitly test which pipeline more closely matches the orbit derived with later epochs of data using a simple cross-validation test. In this, we produce an orbit fit for Weywot with each discrepant data set and compare the fit quality. \trackchange{We find the typical residuals are about an order of magnitude smaller when using our reduction, so we adopt it for the remainder of the analysis.}

In addition to the six HST detections of Weywot in this reanalyzed dataset, two additional archival observations did not detect Weywot due to its proximity to Quaoar. In principle, non-detections can be used to constrain Weywot's orbit, but doing so effectively requires a detailed analysis of the detection characteristics of the instruments used. Since the constraints placed would be relatively uninformative, especially in light of our extensive dataset, we elect to ignore these non-detections. 

\subsection{New Observations}
Since 2023, six new epochs of Hubble Space Telescope observations have been acquired of Quaoar-Weywot using the Wide Field Camera 3 (WFC3). These observations were taken as part of HST Programs 17315 and 17417. PSF fitting of each image with TinyTim \citep{krist201120} model PSFs was completed using the same methods to those used for the archival reanalysis. This method has been extensively validated for WFC3 observations across a variety of TNO binary observations \citep{grundy2015mutual,grundy2019uk126}. We conservatively place a 1 mas noise floor on the astrometry to account for any systematic errors in the astrometric reduction (i.e. time variable distortion of the WFC3 field, uncertainties in pixel scale, etc.). Of the six epochs of WFC3 data, five were able to have astrometry extracted, with the one remaining epoch having an unluckily oriented PSF relative to Weywot which prevented high-quality PSF fitting.


In addition to WFC3 observations, since 2019, Weywot has been detected in four separate stellar occultations \citep{Kretlow2019,fernandez2023weywot,braga2025investigating}. Although a Weywot detection alone does not necessarily provide relative position information, Quaoar's ephemeris is well-known enough that the on-sky offset can be derived. We use the derived relative astrometry for the first three occultations based on circular limb fitting in \citet{braga2025investigating}. In that work, they derive a ``northern'' and ``southern'' solution, which we average into a single data point. For the last occultation, we use the derived position of Weywot from Fernandez-Valenzuela, in prep., which performed an elliptical limb fit to derive Weywot's position. 

For occultation derived astrometry, various modeling choices can slightly influence the measured position of Weywot, like limb shape, timing errors, and many others. To account for these systematic effects, we place a 3 mas noise floor on all occultation derived astrometry, approximately half the angular diameter of Weywot. 

Our entire astrometric dataset is shown in Table \ref{tab:observations}. All HST data used in this paper, including the reanalyzed archival data, was retrieved from the Mikulski Archive for Space Telescopes (MAST) at the Space Telescope Science Institute. The complete set of observations analyzed can be accessed via \dataset[DOI: 10.17909/ex7s-3w78]{http://dx.doi.org/10.17909/ex7s-3w78}.

\section{Orbit Models}
\label{sec:models}
In this work, we fit the entire catalogue of astrometric data using \multimoon, an orbit fitter designed to work with TNO binaries \citep{ragozzine2024beyond}. \multimoon{} treats the orbit fitting as a Bayesian parameter inference problem and uses a Markov Chain Monte Carlo (MCMC) approach. The MCMC sampler is built around \texttt{emcee} \citep{foreman2019emcee}, a popular out-of-the-box ensemble MCMC sampler. \multimoon{} is able to complete both Keplerian and non-Keplerian fits, accounting for the nonspherical shape of Quaoar's gravitational field. In this work, we complete a series of Keplerian and non-Keplerian orbit fits, allowing us to infer the orbital properties of Weywot and constrain the shape and spin axis of Quaoar. For further details on how \multimoon{} works and is used, see \citet{ragozzine2024beyond} and \citet{proudfoot2024bpm2}.

Our initial orbit fitting was focused on finding a Keplerian orbit fit. Under a Keplerian orbit framework, seven free parameters fully describe the orbit of Weywot (system mass, semi-major axis, eccentricity, inclination, and three orbit orientation angles). Our fits allow all seven parameters to vary freely. Orbit fits were run similar to past uses of \multimoon's Keplerian orbit fitter, for more details, see \citet{proudfoot2024bpm2}. 

Based on this set of initial Keplerian orbits fits, we found a relatively poor fit quality ($\chi^2=63.5$, with 29 degrees of freedom), motivating the use of a more complex orbit model. We tested several different orbit models, including non-Keplerian models, models with offsets between Quaoar's center of body (COB) and center of light (COL), and models using an outlier marginalization likelihood function \citep[see][for more details]{proudfoot2024bpm3}. We also tested combinations of these models, finding that the best model to describe our data was a non-Keplerian model with COB-COL offsets. Typically, model comparison requires the use of comparative metrics (such as Bayes factors, bayesian information criterion, etc.), however, since our models are all subsets of each other (e.g., a model with no offsets is equivalent to the offset model where the offsets are set to zero), comparison between models is vastly simplified. \added{We find that including COB-COL offsets is strongly justified when comparing two models with and without COB-COL offsets. The fit with COB-COL offsets produces a $\chi^2 = 17.3$ (0.75 per degree of freedom) while the fit without COB-COL produces $\chi^2 = 44.5$ (2.42 per degree of freedom), giving a likelihood-ratio test statistic of $1.2\times10^{-6}$ \citep[see Section 4 of][]{proudfoot2024bpm2}. Both the improvement in $\chi^2$ per degree of freedom and the likelihood-ratio test indicate a strong preference for the COB-COL model.}

The orbit model for the fits we describe here have 13 free parameters describing Quaoar and Weywot's individual masses, Weywot's six orbital elements, Quaoar's \jt{} and spin pole direction, and two constant offsets between Quaoar's COB and COL in ecliptic latitude and longitude (discussed further below). Non-Keplerian effects were implemented using \multimoon's standard non-Keplerian fitting module, as described in previous work. We also input Quaoar's spin period of $\sim$17.752 hours \citep{kiss2024visible}. The spin period controls Quaoar's axial precession period caused by Weywot's gravitational influence, which is expected to be minimal given Weywot's (presumably) small mass. Our testing in past works shows that the spin period of the primary only has a minor contribution to the overall dynamics, especially with large primary-secondary size differences \citep{proudfoot2024bpm2}.

Our orbit fits neglect \ct{} and all other higher-order harmonics. \ct{} can be safely ignored as its influence is only relevant when the spin-orbit period ratio is close to unity (i.e. near the corotation radius), or near low-order spin-orbit resonances \citep{proudfoot2021prolate}. 
While Quaoar has higher-order gravitational harmonics that may affect the rings, the strength of higher order harmonics fall off at $r^{-3}$ (or greater). At Weywot's distance, they are expected to make only minor contributions to Weywot's dynamics. 

Initial walker positions for all parameters were chosen based on high-likelihood regions identified in preliminary fits. Our orbit fits were run with 960 walkers for 10000 burn-in steps, 5000 post-pruning burn-in steps, and 25000 sampling steps. Convergence was assessed based on autocorrelation times, as well as the smoothness of joint and marginal posterior distributions.

\subsection{COB-COL Offsets}
COL-COB offsets, while described in the appendix of \citet{proudfoot2024bpm2}, have never been fully explored within \multimoon. Orbit fitting for Pluto and Charon made significant use of COB-COL offsets, as both Pluto and Charon have large albedo features that create systematic, time-varying effects in Pluto-Charon astrometry. Surprisingly, COB-COL offsets can not only provide a time-variable signal, but can also cause a systematic offset in relative astrometry. This effect was fully modeled for Pluto and Charon using surface maps derived from resolved HST observations, as well as mutual events \citep{buie1992albedo,1999AJ....117.1063Y}. Unfortunately, no such surface maps of Quaoar are available, forcing us to approximate any COB-COL effects. 

Although Quaoar's true COB-COL offset likely varies in time as a result of rotating albedo features on its surface, a constant offset \added{represents a first-order solution that can correct for \textit{some} of the primary effects of albedo variegations. Longitude-only features (confined between two lines of longitude) albedo features will provide a periodic signal over a single rotation period that provides no systematic COB-COL offset. Latitude-only (between two lines of latitude) features on the other hand, provide a constant COB-COL offset over Quaoar's rotation period. For features with both longitudinal and latitudinal components, effects can be both periodic and systematic. However, with a \textit{large} number of samples,  COB-COL variations will tend to average into an effective constant offset---though not necessarily a perfect one, and typically with a systematically higher $\chi^2$ than expected. Of course, this averaging assumes that our data uniformly sample across Quaoar’s longitude. As we discuss in Section \ref{sec:systematics}, our data set may not uniformly sample longitudes, which may limit the extent of the longitudinal averaging in practice. Unfortunately, implementing a rotationally variable COB-COL is quite difficult, }\trackchange{given the lack of a precise rotational period solution for Quaoar \citep[compare][]{ortiz2003rotational,kiss2024visible,margoti2024quaoarshape}. \added{At this early stage, with the simplicity of} implementing a constant offset, we defer a more physically realistic model to future work. }

\added{Although simplistic, this model allows for exploration of a range of features that were seen on Pluto and Charon. Charon has a $\sim$longitudinally-constant dark south pole, while Pluto has a dark band of red material. Such features provide a systematic effect on $\sim$all observations, which can be accounted for with our constant COB-COL offset.}


With Quaoar's motion around the Sun, its pole orientation slowly varies, which could make a constant offset a poor approximation of latitudinal albedo features. However, at Quaoar's current orbital position, a constant offset in ecliptic latitude and longitude provides an excellent approximation of the changing pole orientation (as shown in Figure \ref{fig:col-cob}). 

We note that our occultation-derived relative astrometry is referenced to Quaoar's ephemeris center, rather than the true position of Quaoar during each occultation. Quaoar's ephemeris is conditioned upon telescopic observations with similar systematic COB-COL offsets as resolved imaging. As such, we treat occultation astrometry in a similar manner to all other resolved observations when fitting COB-COL offsets. In the future, an entire dataset made of double occultations (where both Quaoar and Weywot are detected) could obviate the need for COB-COL offsets. 

\begin{deluxetable*}{lc|ll|ll}
\tablecaption{Non-Keplerian Orbit Solution for Weywot}
\tablehead{
Parameter &  & \text{Constrained fit:} & $\chi^2=37.2$ & \text{Unconstrained fit:} & $\chi^2=17.3$
}
\startdata
\textbf{\textit{Fitted parameters}}         &                  & \textbf{\textit{Priors}}                                               &                                & \textbf{\textit{Priors}}                                               &                                  \\
Mass, Quaoar ($10^{18}$ kg)                 & $M_{q}$          & $\mathcal{U}\left(0,10^{6}\right]$                                     & $1212.27^{+5.39}_{-5.39}$      & $\mathcal{U}\left(0,10^{6}\right]$                                     & $1209.35^{+5.74}_{-5.58}$        \\
Mass, Weywot ($10^{18}$ kg)                 & $M_w$            & $\mathcal{N}\left(2.4,1.2\right)$ & $2.4^{+1.2}_{-1.1}$            & $\mathcal{N}\left(2.4,1.2\right)$ & $2.5^{+1.2}_{-1.2}$              \\
Semi-major axis (km)                        & $a$              & $\mathcal{U}\left[500,10^{10}\right]$                                  & $13329^{+19}_{-19}$            & $\mathcal{U}\left[500,10^{10}\right]$                                  & $13334^{+20}_{-20}$              \\
Eccentricity                                & $e$              & $\mathcal{U}\left[0,1\right)$                                          & $0.0111^{+0.0044}_{-0.0040}$    & $\mathcal{U}\left[0,1\right)$                                          & $0.0097^{+0.0027}_{-0.0024}$     \\
Inclination ($\degr$)                       & $i$              & $\mathcal{U}\left[0,180\right]$                                        & $13.62^{+0.32}_{-0.33}$        & $\mathcal{U}\left[0,180\right]$                                        & $14.83^{+0.44}_{-0.45}$          \\
Apsidal argument ($\degr$)                  & $\omega$         & $\mathcal{U}\left[0,360\right]$                                        & $97^{+28}_{-37}$               & $\mathcal{U}\left[0,360\right]$                                        & $298^{+56}_{-47}$                \\
Nodal longitude ($\degr$)                   & $\Omega$         & $\mathcal{U}\left[0,360\right]$                                        & $353.3^{+0.83}_{-0.85}$        & $\mathcal{U}\left[0,360\right]$                                        & $353.07^{+1.03}_{-1.25}$         \\
Mean anomaly ($\degr$)                      & $\mathcal{M}$    & $\mathcal{U}\left[0,360\right]$                                        & $79^{+39}_{-27}$               & $\mathcal{U}\left[0,360\right]$                                        & $237^{+47}_{-56}$                \\
$J_2$ harmonic, Quaoar                      & $J_2$            & $\mathcal{U}\left[0,10\right]$                                         & $0.018^{+0.009}_{-0.008}$      & $\mathcal{U}\left[0,10\right]$                                         & $0.379^{+0.160}_{-0.191}$        \\
Axis obliquity, Quaoar ($\degr$)            & $i_{sp}$         & $\mathcal{N}\left(12.6,3.0\right)$                                     & $16.53^{+1.93}_{-1.78}$        & $\mathcal{U}\left[0,180\right]$                                        & $14.72^{+0.34}_{-0.28}$          \\
Axis precession, Quaoar ($\degr$)           & $\Omega_{sp}$    & $\mathcal{N}\left(331.46,3.0\right)$                                   & $332.84^{+3.06}_{-3.03}$       & $\mathcal{U}\left[0,360\right]$                                        & $353.48^{+0.78}_{-1.08}$         \\
COL longitude offset (mas)                  & $\Delta x_{col}$ & $\mathcal{U}\left[-20,20\right]$                                       & $0.3^{+3.0}_{-2.9}$              & $\mathcal{U}\left[-20,20\right]$                                       & $-0.8^{+3.0}_{-2.4}$               \\
COL latitude offset (mas)                   & $\Delta y_{col}$ & $\mathcal{U}\left[-20,20\right]$                                       & $5.5^{+1.4}_{-1.4}$            & $\mathcal{U}\left[-20,20\right]$                                       & $4.8^{+0.9}_{-0.9}$              \\
\hline
\textbf{\textit{Derived parameters}}        &                  & \textbf{\textit{Priors}}                                               &                                & \textbf{\textit{Priors}}                                               &                                  \\
Keplerian orbital period (d)                & $P_{orb}$        & \nodata                                                                & $12.42727^{+0.00003}_{-0.00003}$ & \nodata                                                                & $12.42718^{+0.00024}_{-0.00009}$ \\
Quaoar-centric inclination ($\degr$)        & $\varepsilon$    & \nodata                                                                & $6.13^{+1.35}_{-1.06}$         & \nodata                                                                & $0.41^{+0.32}_{-0.21}$           \\
Orbit pole RA ($\degr$)                     & $\alpha_{w}$     & \nodata                                                                & $267.38^{+0.33}_{-0.32}$                        & \nodata                                                                & $267.15^{+0.44}_{-0.52}$                          \\
Orbit pole dec. ($\degr$)                   & $\delta_{w}$     & \nodata                                                                & $53.00^{+0.33}_{-0.32}$                        & \nodata                                                                & $51.80^{+0.45}_{-0.43}$                          \\
Mean motion ($\degr$ d$^{-1}$)              & $n$              & \nodata                                                                & $28.9586^{+0.0006}_{-0.0006}$  & \nodata                                                                & $28.93^{+0.01}_{-0.01}$          \\
Apsidal precession rate ($\degr$ yr$^{-1}$) & $\dot{\varpi}$   & \nodata                                                                & $0.48^{+0.24}_{-0.21}$         & \nodata                                                                & $10.19^{+4.29}_{-5.13}$          \\
Nodal precession rate ($\degr$ yr$^{-1}$)   & $\dot{\Omega}$   & \nodata                                                                & $-0.49^{+0.21}_{-0.24}$        & \nodata                                                                & $-10.19^{+5.13}_{-4.29}$      \\  
\hline
\enddata
\tablecomments{Reported values represent the median value and uncertainties are based on 16th and 84th percentile values. Priors are either uniform ($\mathcal{U}$) with upper and lower bounds listed, or normally distributed ($\mathcal{N}$) with mean and scale listed. All fitted angles are relative to the J2000 ecliptic plane on Quaoar-centric JD 2454000.0 (2006 Sep 21 12:00 UT), except for RA and dec. values which are referenced to the J2000 equatorial coordinate system. Assumed c-axis for Quaoar is 545 km and spin period is 17.752 hours \citep[][]{kiss2024visible}, although realistic variation of these values produces no meaningful change to the fit. To transform to $J_2$ from the more physically meaningful $J_2R^2$, we use a volumetric radius of 549 km.}
\label{tab:fits}
\end{deluxetable*}

\subsection{Priors}
Priors for most parameters were chosen to be uninformative, mostly limiting parameters to physical/realistic values. For Weywot's mass, which is relatively unconstrained by our orbit model, we placed a Gaussian prior with mean of $2.4\times10^{18}$ kg and standard deviation of $1.2\times10^{18}$ kg, roughly corresponding to the mass expected with a diameter of 165 km (E. Fernández-Valenzuela, in prep.) and density between 500-1500 kg m$^{-3}$. Since Weywot's mass is very weakly correlated with Quaoar's \jt, this ensures the most physically realistic model, which is important when later characterizing the dynamics of the ring system. 

Quaoar's spin pole direction is constrained by the pole orientation of its two rings. As collisional systems, rings tend to damp to the local Laplace plane (i.e. the plane that minimizes perturbations), which is usually close to a body's equatorial plane. Based on the precession rates from Quaoar's \jt{} and Weywot, the Laplace plane should be $\lesssim0.5\degr$ inclined from Quaoar's equatorial plane (assuming a large range of Weywot densities). Thus, we can assume Quaoar's pole orientation is close to that of the ring plane. Recent work combining all detections of Quaoar's rings---including new detections with JWST---have enabled precise measurement of Quaoar's ring pole \citep{proudfoot2025jwst}. This work is the first to use a self-consistent model of the rings which accounts for their changing geometry since their first detection in 2018. We use those measurements as a Gaussian prior, although we inflate the uncertainties to 3$\degr$. This accounts for any systematic effects in the ring modeling procedures. For example, Q1R is expected to have a fairly high eccentricity \citep{rodriguez2023dynamical}, which can bias measurements of pole direction when modeling the rings as circular. Likewise, any warps or bending waves (see Section \ref{sec:rings}) may also bias ring pole determinations. For comparison, we also run a fit without any priors on Quaoar's pole orientation. We refer to these two different fits as the constrained fit and the unconstrained fit throughout the rest of this paper\footnote{\added{We note that the COB-COL likelihood ratio test for the constrained fit produces a likelihood ratio of $8.3\times10^{-6}$, still indicating a strong preference for a model with the COB-COL offset included}.}.



\begin{figure*}
    \centering\includegraphics[width=0.99\textwidth]{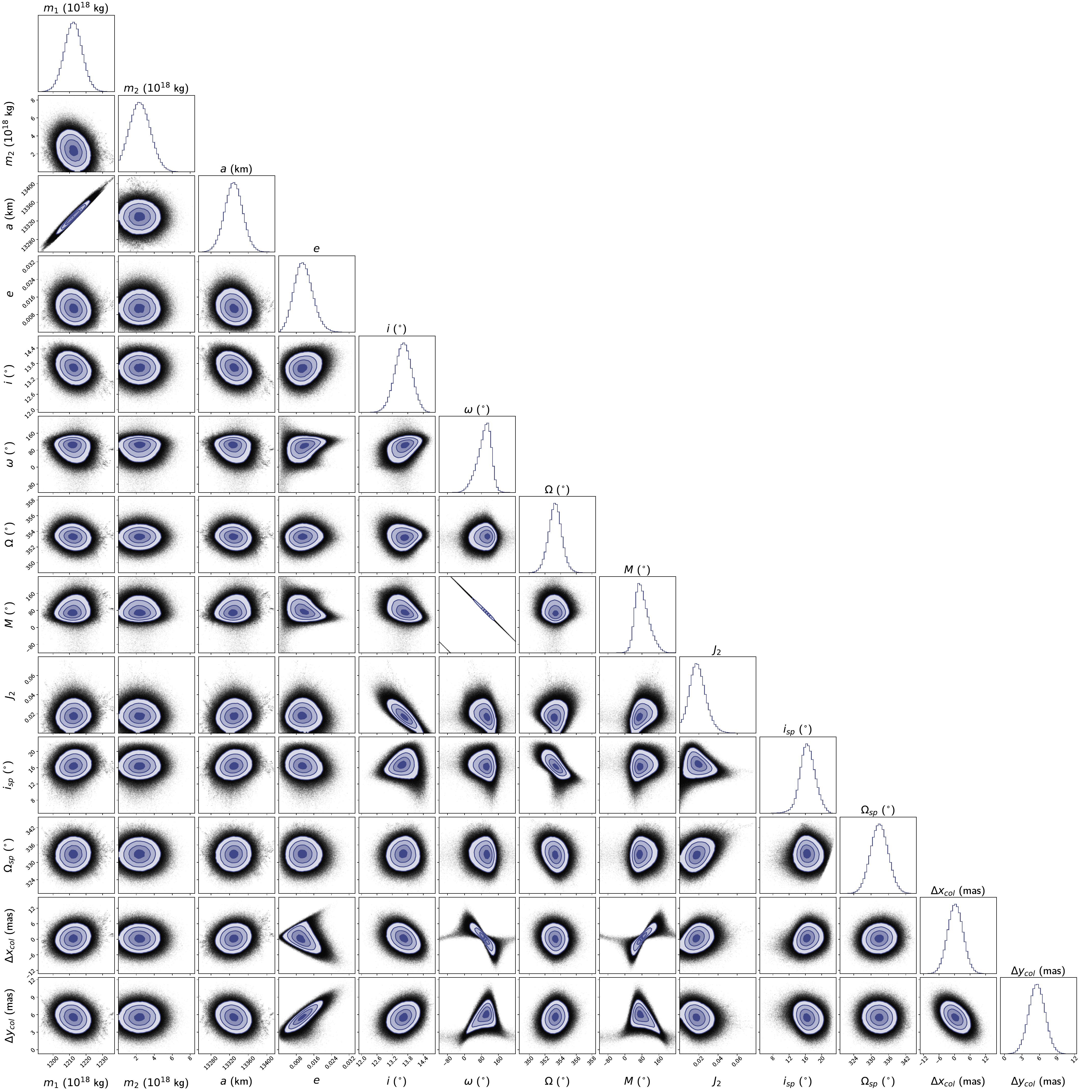}
    \caption{A corner plot showing the posterior distribution from our constrained orbit fit. Along the diagonal are the marginal (1D) posterior distributions for each parameter, alongside the 2D joint posterior distributions for each pair of parameters. Contours on the joint distributions show the 1$\sigma$, 2$\sigma$, and 3$\sigma$ confidence regions, and black points correspond to individual samples from the MCMC chain.}
    \label{fig:corner_con}
\end{figure*}

\begin{figure*}
    \centering\includegraphics[width=0.99\textwidth]{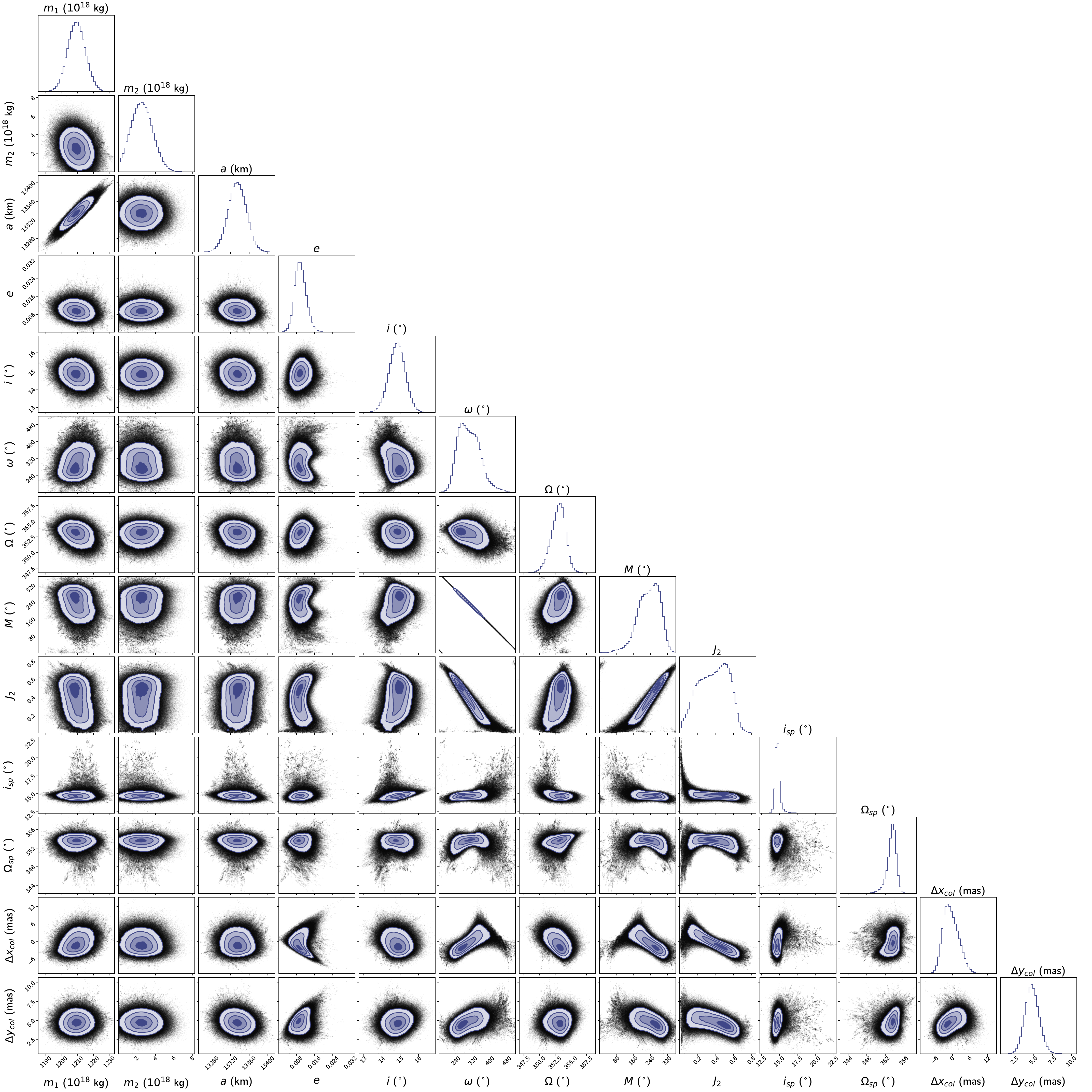}
    \caption{Similar to Figure \ref{fig:corner_con}, but for the unconstrained orbit fit.}
    \label{fig:corner_uncon}
\end{figure*}

\section{Orbit Fitting Results}
\label{sec:results}

The results of our orbit fits are presented in Table \ref{tab:fits} alongside a variety of derived parameters. We also show corner plots of the posterior distributions in Figures \ref{fig:corner_con} and \ref{fig:corner_uncon}. These orbit fits represent the most detailed exploration of Weywot's orbit to date.

\begin{figure}
    \includegraphics[width=\linewidth]{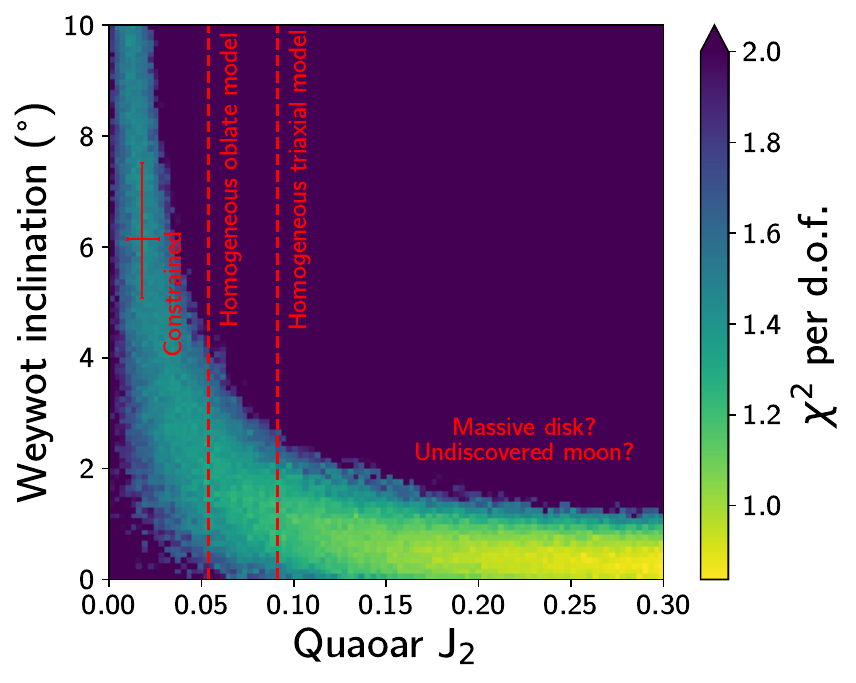}
    \caption{Orbit fit quality ($\chi^2$ per degree of freedom) as a function of Quaoar's \jt{} and Weywot's inclination (w.r.t. Quaoar's equatorial plane). Red dashed lines show the values expected from the undifferentiated oblate and triaxial shape models. The red cross shows the \jt{} and inclination value measured in the constrained orbit fit. We note that the best fit quality is at $J_2 \sim 0.4$, but we limit the range of \jt{} displayed to better show the fit quality at more realistic values of Quaoar's \jt. For reference, the p-value of a $\chi^2_{pdf}=1$ (1.8) is 0.46 (0.01).}
    \label{fig:j2-inc}
\end{figure}

Comparison between the orbit fits show many fundamental differences. Most importantly, the fit quality of the unconstrained orbit fits is substantially better, with a best fit $\chi^2$ per degree of freedom of 0.80 compared to 1.62 for the constrained fit. This difference in fit quality is primarily a function of Quaoar's \jt{} and pole direction. This can be further simplified as a correlation between fit quality, Quaoar's \jt, and Weywot's inclination (which simplifies the complex dependency between orbit and spin orientation angles). Fit quality is relatively high at low Weywot inclination, with a broad range of allowable values for Quaoar's \jt. Conversely, at high Weywot inclination ($\gtrsim$1$\degr$), fit quality is much lower, but permits a small range of non-zero values for Quaoar's \jt. We show this correlation in Figure \ref{fig:j2-inc}. 

One possible explanation of the large difference in fit quality between the orbit fits is systematic errors induced by time-varying COB-COL offsets, which are overfitted by the unconstrained model. As an example, consider the astrometry of Weywot when Quaoar has a time varying COB-COL offset. The motion of the COL will be some complicated periodic function set by the albedo distribution, Quaoar's rotation period, Quaoar and Earth's heliocentric motion, the telescope filter (since the albedo distribution can differ at varying wavelengths), and a variety of other factors. With enough observations, the offset in the direction perpendicular to the pole position angle will tend to average out, given Quaoar's non-synchronous rotation. But given just a handful of observations which are sparsely sampled, the effects of this COB-COL wobble (when unaccounted for) will reduce fit quality. Our new HST observations are quite precise, with uncertainties on the order of 1 mas. With angular diameter of Quaoar of $\sim$40 mas, even small albedo variations could induce significant systematics. Rapid precession on a near circular/equatorial orbit (as seen in our unconstrained fit) would induce similar wobbling motion, which could coincidentally produce a better fit to the astrometry. 

The implied \jt{} in our unconstrained orbit fit is likely to be unrealistically high, unless undiscovered moons remain undetected within Weywot's orbit or Quaoar's ring system is unexpectedly massive. Likewise, the pole orientation of Quaoar in these fits is very different than that implied by the orientation of Quaoar's ring system, with a $\sim5\degr$ inclination. In comparison, the constrained orbit fit is roughly consistent with shape/interior models for a differentiated Quaoar (see Section \ref{sec:interior} for more exploration of this). Given the constrained fits seem to be more physically realistic in terms of Quaoar's \jt{} and pole orientation, we believe the constrained model likely provides a more accurate representation of the Quaoar-Weywot system, even with the lower fit quality. 

Other than the differences in Quaoar's \jt---and the corresponding changes in orbit/spin orientation angles---our two orbit fits are remarkably consistent with one another. Both fits provide a determination of Quaoar's mass (when using the Weywot mass prior), with only 0.5\% uncertainties in both orbit fits. Likewise, the orbit fits provide precise determinations of Weywot's semi-major axis, which is important for understanding the location of Weywot's 6:1 MMR near the outer ring. Combined, the mass and semi-major axis determinations allow for an ultra-precise measurement of Weywot's orbit period, with uncertainty of $<$20 seconds---and 10$\times$ more precise in the constrained fit. 

\begin{figure*}
    \centering\includegraphics[width=0.6\textwidth]{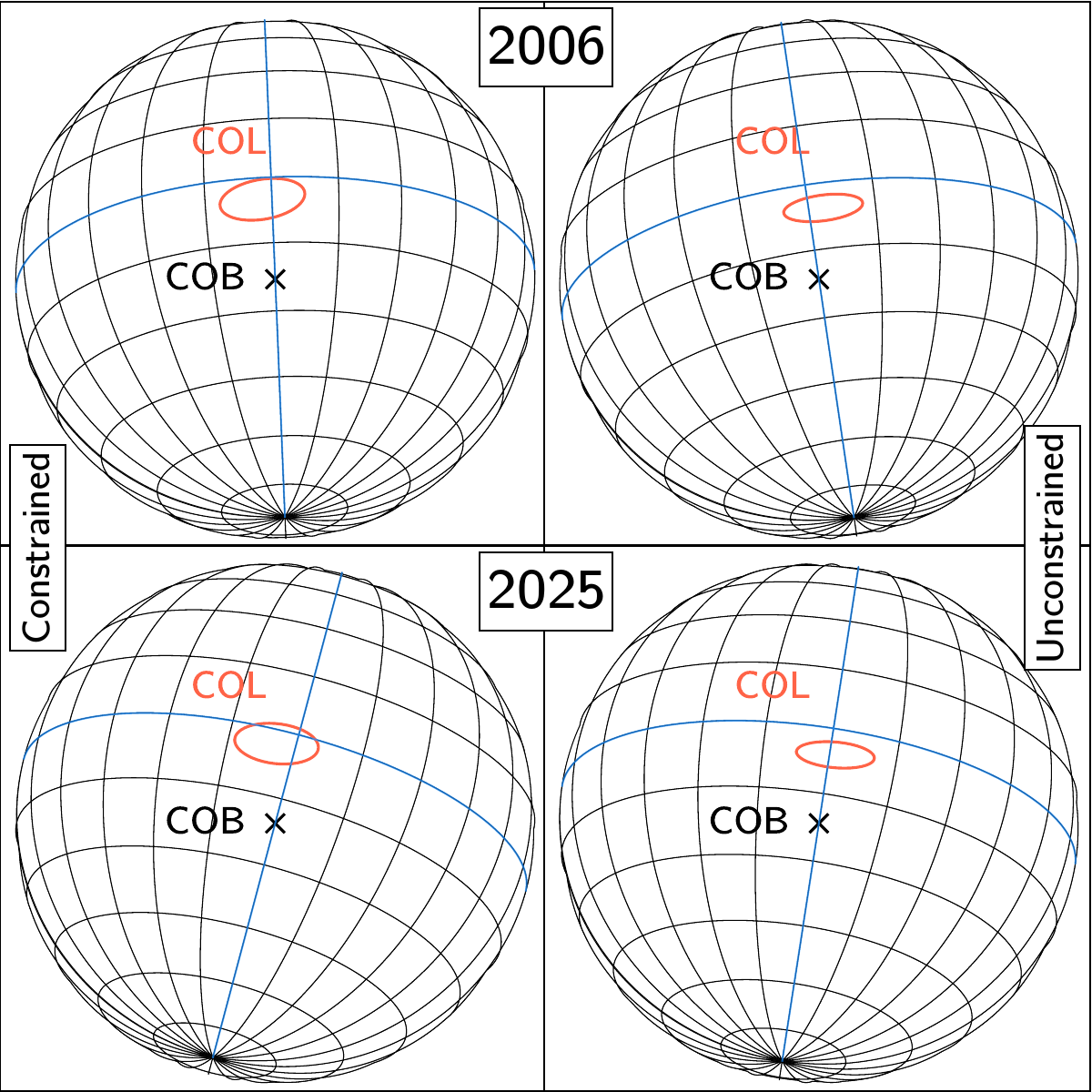}
    \caption{The COL-COB offsets implied by our constrained (left) and unconstrained (right) orbit fits. The COL is shown as a red ellipse (based on the orbit fit posterior) and the COB is marked with a black x. Quaoar is shown as a sphere with with angular diameter of 36 mas. The equator and sub-observer longitude are shown with blue lines. Over our observational baseline, a constant offset in ecliptic latitude and longitude is a good approximation of a COB-COL offset caused by longitudinal albedo features.}
    \label{fig:col-cob}
\end{figure*}

Both our orbit fits also show a small eccentricity, \trackchange{differing from most previous orbit fits in the literature. This is largely the result of our astrometric reanalysis described in Section \ref{sec:reanalysis}.} We consistently find $e<0.02$, even without inclusion of any data from 2019 or later. Recently published work using our reanalyzed dataset---but not our new HST data---shows a similarly low eccentricity, emphasizing the importance of \trackchange{the} reanalysis \citep{braga2025investigating}. 

In our fits, $e$ is also strongly correlated with the COB-COL offsets. To first order, a circular orbit with a small COB-COL offset appears as a slightly eccentric orbit, so seeing such a correlation is not unusual \citep[e.g.,][]{buie2012orbit}. More realistic COB-COL offsets should further refine Weywot's actual eccentricity, but here, we place an upper limit of $e<0.02$. Weywot's eccentricity is important for understanding the dynamics of Quaoar's rings, and this upper limit limits their dynamical strength. 

One of the most interesting aspects of our orbit fits is the strong preference for non-zero COB-COL offsets. We find strong evidence for COB-COL offsets, with both orbit fits measuring a non-zero average offset at $\gtrsim$4$\sigma$ confidence. Preliminary Keplerian orbit fits also found non-zero offsets at $\sim4.5\sigma$ confidence. As discussed previously, these offsets are not physically realistic by construction; however as can be seen in Figure \ref{fig:col-cob}, the constant offset (in ecliptic lat. and lon.) tracks similar positions on Quaoar's surface throughout our observational baseline. This coincidentally allows the constant COB-COL offset to provide a physically meaningful model, validating our method of implementation. 

In both fits (and preliminary Keplerian fits), the offset implied is roughly aligned with the projected pole direction of Quaoar, which is consistent with significant latitudinal albedo variations across Quaoar's surface. Although this could occur randomly by chance ($\sim5\%$), this alignment is suggestive that the COB-COL offsets in our models represent real albedo features. Interestingly, for both our orbit fits, the implied COL (shown as a red ellipse in Figure \ref{fig:col-cob}) is nearly coincident with Quaoar's equator. This could provide a hint into the albedo distribution of Quaoar. In Section \ref{sec:offsets}, we discuss this further. 

Given the long observational baseline of our data, we are able to fully determine the orbital plane of Weywot. With a small dataset, orbit fitting is generally not able to fully determine a satellite's orbital plane, leaving two ambiguous solutions that are mirror images of one another. \trackchange{Mirror solutions have never been formally ruled out in the literature, making it unclear if mirror solutions have ever been formally excluded.} During our early Keplerian orbit fits, we found that the mirror solutions provided a much worse fit to the data and could be excluded at $>$50$\sigma$ confidence.

To facilitate future observations of Weywot, predictions of occultations, and a variety of other scientific investigations, we include an ephemeris of Weywot referenced to Quaoar's COL. This ephemeris---spanning from 2005 until 2035---is generated by taking 500 random posterior samples from the constrained fit MCMC chain. The first 10 rows of this ephemeris are displayed in Table \ref{tab:ephem}, with the rest of the ephemeris available as a machine readable table in the online version of this article. 

\begin{deluxetable*}{ccCCCCCC}
\tablecaption{System Ephemeris}
\tablehead{
Julian Date & Date & \Delta \alpha \cos{\delta} & \sigma_{\Delta \alpha \cos{\delta}} & \Delta \delta & \sigma_{\Delta \delta} & r & \sigma_{r} \\ 
& & ('') & ('') & ('') & ('') & ('') & ('')}
\startdata
2453371.5  & 2005-01-01 0:00:00  & 0.36355 & 0.00124 & -0.11581 & 0.00137 & 0.38152 & 0.00129 \\
2453371.75 & 2005-01-01 6:00:00  & 0.38515 & 0.00109 & -0.10207 & 0.00128 & 0.39845 & 0.00111 \\
2453372    & 2005-01-01 12:00:00 & 0.40074 & 0.00100 & -0.08664 & 0.00124 & 0.41003 & 0.00100 \\
2453372.25 & 2005-01-01 18:00:00 & 0.41003 & 0.00098 & -0.06977 & 0.00124 & 0.41591 & 0.00098 \\
2453372.5  & 2005-01-02 0:00:00  & 0.41287 & 0.00104 & -0.05170 & 0.00129 & 0.41610 & 0.00105 \\
2453372.75 & 2005-01-02 6:00:00  & 0.40918 & 0.00118 & -0.03272 & 0.00139 & 0.41053 & 0.00119 \\
2453373    & 2005-01-02 12:00:00 & 0.39901 & 0.00140 & -0.01314 & 0.00150 & 0.39926 & 0.00141 \\
2453373.25 & 2005-01-02 18:00:00 & 0.38248 & 0.00168 & +0.00674  & 0.00164 & 0.38261 & 0.00167 \\
2453373.5  & 2005-01-03 0:00:00  & 0.35985 & 0.00200 & +0.02660  & 0.00178 & 0.36089 & 0.00197 \\
2453373.75 & 2005-01-03 6:00:00  & 0.33144 & 0.00233 & +0.04612  & 0.00193 & 0.33472 & 0.00229 \\
\hline
\enddata
\tablecomments{The predicted right ascension and declination positions of Weywot---referenced to Quaoar's COL---from 2005 through 2035. Predicted positions, separations, and uncertainties are taken from a sample of 500 random posterior draws from the constrained orbit fit. We display the first 10 rows of the table with the rest of the table available as a machine-readable table.}
\label{tab:ephem}
\end{deluxetable*}

\begin{figure*}
    \centering\includegraphics[width=0.99\textwidth]{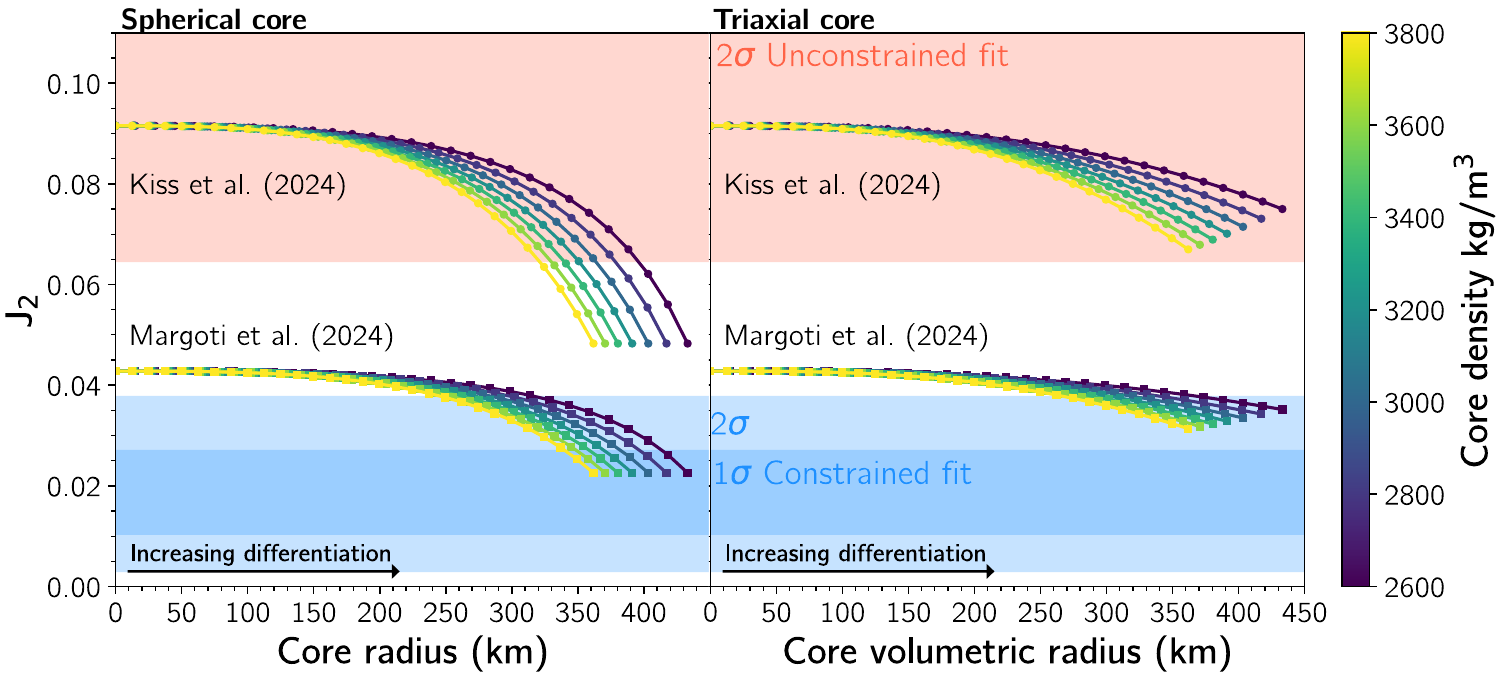}
    \caption{\jt{} values for Quaoar as a function of core size and core density assuming a two-layer interior structure model with spherical cores (left) and triaxial cores (right). The two-layer interior model consists of a rocky core (with chosen core density) surrounded by a partially differentiated shell of mixed ice and rock. Points at core radii of 0 km correspond to an undifferentiated model, and the maximum core radii corresponds to a fully differentiated body with shell density of water ice (920 kg m$^{-3}$). Core densities are chosen to bound likely core densities of TNOs with 2600 kg m$^{-3}$ corresponding to hydrated rock and 3800 kg m$^{-3}$ corresponding to dehydrated rock. }
    \label{fig:interior}
\end{figure*}

\section{Quaoar's \jt{} and interior}
\label{sec:interior}

\subsection{Shape Models}
As guides for our interpretation of Quaoar's \jt, we consider two recently published shape models for Quaoar. The first shape model was developed to fit the visible and thermal light curves of Quaoar, as well as its overall thermal emission \citep{kiss2024visible}. The thermophysical study, assuming a constant global albedo, predicted axes ratios of $a/b = 1.19$ and $b/c = 1.16$, with a range of possible volumetric radii. We assume a volumetric radius of 549 km (to simplify comparison between shape models), which gives $a = 648$ km, $b = 544$ km, and $c = 469$ km. We refer to this as the light curve shape model.

In contrast, \citet{margoti2024quaoarshape} used 62 positive occultations chords over 20 different occultations to derive the three-dimensional shape of Quaoar. This method allows a better determination of the shape, without requiring any assumptions about Quaoar's albedo, giving $a = 583.3\pm2.6$ km, $b = 555.3\pm1.0$ km, and $c = 510.0\pm1.0$ km, and a volumetric radius of $548.8\pm1.1$ km. We refer to this as the occultation shape model.

With both of these shape models, we can derive Quaoar's expected \jt{} assuming various internal density structures. With a homogeneous interior (i.e. undifferentiated), \jt{} is a function of the semi-axes $a$, $b$, $c$, where
\begin{equation}
    J_2 = \frac{1}{10R^2}\left(a^2 + b^2-2c^2\right)
\end{equation}
\noindent \citep{1995geph.conf....1Y}, giving $J_2=0.092$ and $J_2 = 0.043$ for the light curve and occultation models, respectively. Quaoar, however, as a large TNO has likely sustained high temperatures in its interior, allowing at least partial differentiation. Its large density hints that gravitational compaction, or even melting, could have eliminated internal pore space \citep{bierson2019using}. This is further evidenced by light hydrocarbons on its surface, which could be derived from internal geochemical processes \citep{emery2024tale}. 

\subsection{How does differentiation change \jt?}
To account for differentiation, we can instead consider a two layer density model, with a spherical rocky core at its center surrounded by a partially (or totally) differentiated shell. Selecting a core radius and density, we can calculate a core volume and mass, which sets the shell shape, volume, and density (based on the shape model and mass given by our orbit fits). The gravitational harmonics of a given shape (with mass, $M$) are determined by its principal moments of inertia. The \jt{} is given by
\begin{equation}
\label{eqn:moments}
    J_2 = \frac{C - \frac{1}{2} \left( B + A \right)}{MR^2}
\end{equation}
\noindent \citep{1995geph.conf....1Y}, where $A,B,C$ are the principal moments of inertia around the $a,b,c$ axes. Using the shape and density of the core and shell, we can calculate the \jt{} solely as a function of the shape model, core density, and core radius. We show these relationships in Figure \ref{fig:interior}, where the far left models correspond to an undifferentiated body, and the far right models correspond to a fully differentiated body with a shell density of pure water ice (920 kg m$^{-3}$). We select a range of core densities based on the densities of hydrated and dehydrated rock, 2600 kg m$^{-3}$ and 3800 kg m$^{-3}$, respectively \citep{dunham2019haumea,noviello2022let}. Assuming a spherical core places a lower bound on the \jt{} when considering differentiation, since Quaoar's core is likely not spherical, especially given Quaoar's external shape.

To more accurately consider core shape, we also complete the same analysis as above, but instead model the core with the same axes ratios as the external shape. We show this sequence of models on the right panel of Figure \ref{fig:interior}. Since a rocky core is more dense than the external shell, the core is likely to be somewhat less oblate/prolate than the external figure. Hence, these models are probably more of an overestimate of the \jt, which we consider as a rough upper bound. 

We note that these two-layer models are not fully self-consistent models that satisfy hydrostatic equilibrium, although they serve as a useful guide for understanding how differentation affects \jt. In the future, self-consistent hydrostatic equilibrium models, like those implemented in \texttt{kyushu} \citep{dunham2019haumea}, can be used to jointly interpret Quaoar's \jt{} and external figure. 

\subsection{Quaoar's interior from orbit fits}
With realistic estimates/bounds on possible \jt{} values, we can compare with the results of our orbit fitting to glean insights into the interior of Quaoar. Focusing on the constrained orbit fit, we find that the measured \jt{} is consistent with the differentiated occultation model (when taking the average of the spherical and triaxial core models), but inconsistent with the undifferentiated model. It is also wholly inconsistent with the light curve model, even when considering differentiation. Looking to the unconstrained fit, both shape models provide poor fits, although the undifferentiated light curve shape model fits the \jt{} posterior at $\sim$2$\sigma$. 

Assuming that the constrained fit is a better description of Weywot's orbit, we suggest that Quaoar is differentiated and has a shape consistent with the occultation-derived shape model. As this model predicts a lower amplitude light curve than observed \citep{margoti2024quaoarshape}, we further suggest that Quaoar's light curve is shaped by albedo features. Based on our measurement of Quaoar's mass, the occultation shape model produces a density of $1751\pm13$ kg m$^{-3}$. We note, however, at such precision, systematics from other effects (such as topography, slight non-triaxiality, etc.) could be important for a precise measurement of Quaoar's bulk density. 

During our preliminary fits, we also explored orbit models with robust outlier detection methods implemented \citep[as in][]{proudfoot2024bpm3}. These models provide a useful way to examine whether our results could be biased by potential outliers or underestimated errors. Those orbit fits show nearly equivalent orbit solutions, although they can only place an upper limit of $J_2<0.026$ (0.042) at $1\sigma$ ($2\sigma$) confidence, fairly similar to the 1/2$\sigma$ limits in the constrained fit. This suggests that our inferences are robust to possible outliers or data quality issues.



We reemphasize that, as seen in Figure \ref{fig:j2-inc}, \jt{} and Weywot's inclination (w.r.t. Quaoar's equator) are strongly correlated. Our constrained fit assumes that Quaoar's pole is aligned with the ring system and that the rings' pole orientation is well-measured. Unlike Chariklo or Haumea, the orientation of Quaoar's rings have \textit{never} been measured at a single epoch. Instead, the orientation of the rings is measured by combining detections from many epochs, which could introduce systematic errors. For example, \citet{rodriguez2023dynamical} suggest that based on width variations in Q1R, the ring is expected to have $e\gtrsim0.036$. Such a ring would slowly precess, creating complications in combining temporally separated detections. This could easily lead to significant uncertainties in our knowledge of the ring pole. As such, \trackchange{fits should} be updated once the ring system is better characterized. We especially encourage large occultation campaigns that could characterize the ring orientation and eccentricity at a single epoch.

\subsection{Massive disks and/or undiscovered satellites?}

Although we do consider the constrained model to provide a better description of Weywot's dynamics, it is worth exploring the implications of the high \jt{} value found in the unconstrained model. Although the unconstrained model is 2$\sigma$ consistent with the triaxial model, the bulk of the posterior has much higher values for \jt. One possible reason for this would be orbiting mass between Quaoar and Weywot. 

Although no definitive measurements of the mass of Quaoar's ring system have been made, we can place an upper limit on it by comparing to Haumea and Chariklo's rings. \citet{sicardy2020dynamics} suggests that Haumea and Chariklo's ring systems likely have similar surface densities as Saturn's A ring, $\sim$500-1000 kg m$^{-2}$. Generously assuming the same surface density, despite it being the least dense small body ring system, we find an upper limit on the mass of Quaoar's rings of $M_r\sim10^{16}$ kg, equivalent to an icy body with a radius of 15 km. The \jt{} contribution of a ring can be derived from Equation \ref{eqn:moments}, giving
\begin{equation}\label{eqn:j2_ring}
    J_{2} \approx \frac{1}{2} \frac{M_{r}}{M_q} \left(\frac{r}{R}\right)^2
\end{equation}
\noindent where $r$ is the ring radius. This yields $J_{2,ring} \sim 0.0002$, contributing almost no detectable signature when compared to our estimates of Quaoar's \jt. Evidently, Quaoar's rings contribute little to the dynamics of Weywot. 

Another possibility is the presence of undetected moons between Quaoar and Weywot. Moons embedded within/near Quaoar's rings would have a maximum angular separation from Quaoar of $\sim$0.15 arcseconds, which is equivalent to $<$4 WFC3 pixels. This separation is difficult to resolve with current telescopes, even JWST. As such, it remains possible for inner satellites to remain undiscovered around Quaoar. To explain the entirety of the \jt{} measured in our unconstrained fit using Equation \ref{eqn:j2_ring}, a satellite orbiting at 5000 km would need a mass of $\sim10^{19}$, roughly equivalent to an icy body with a radius of $\sim$140 km. Such a massive satellite could present stability issues for the ring system, unless the rings are at/near resonances with this putative moon. A system of shepherd satellites could also play a role in raising the apparent \jt, but would require $\sim$thousands of 10 km bodies to generate the same \jt{} signature. 

Another satellite interior to Weywot could help to explain Weywot's substantial inclination relative to the ring plane ($\sim$5$\degr$ in the unconstrained fit). Similar to Hi'iaka and Namaka, Weywot could have its inclination excited by a massive companion \citep{rb09,proudfoot2024bpm3}. 

Although we encourage observations to search for additional satellites around Quaoar, we believe it is unlikely that enough orbiting mass exists around Quaoar to generate the \jt{} signal that our fits imply. As such, we consider the constrained orbit fit to be the more realistic fit, especially when considering its alignment with the ring system.

\section{COB-COL Offsets and Quaoar's Surface}
\label{sec:offsets}

Until now, COB-COL offsets have not been a necessary part of any orbital model for TNO binaries/satellites, with the exception of Pluto. With an angular diameter of up to 110 mas, along with drastic albedo features, Pluto's COB-COL offsets are a necessary part of any orbital model. These offsets were studied at length in the decades before the \textit{New Horizons'} flyby. Surface albedo maps were originally made by taking advantage of Pluto's 1985-1990 mutual event season \citep{buie1992albedo}, which were further supplemented by imaging of Pluto and Charon with HST's ACS instrument that could partially resolve Pluto's surface \citep{buie2010pluto}. Orbit fits without these albedo maps consistently generate non-zero eccentricities and large residuals \citep[e.g.,][]{tholen1997orbit}. One important aspect of these COB-COL offsets is a systematic offset in the astrometry, in addition to a periodic signal phased to the rotation period. This is seen in Figure 9 of \citet{buie2012orbit}, where orbit fits for Pluto and Charon have systematic offsets of $\sim7-8$ mas. With measurement uncertainties of order $\sim$1 mas (see Table \ref{tab:observations}), it quickly becomes clear Quaoar could be significantly affected by similar issues.

In both our orbit fits, we find significant detections of COB-COL offsets. Our constrained fit shows a total COB-COL offset of 5.5$\pm$1.4 mas and our unconstrained fit shows an offset of 4.9$\pm$1.0 mas, corresponding to about 170 km and 150 km at Quaoar's typical distance of $\sim$42 au. These offsets appear to be well-aligned with the projected position angle of Quaoar's pole, suggesting the offsets are modeling a real phenomenon. 

Although large, plausible surface albedo features could explain these offsets. We explored a few toy models of Quaoar's albedo and found that latitudinal variations of albedo (by a factor of a few) could explain these offsets. For example, a map where Quaoar has a bright equatorial band---putatively created by slow infalling of ring material---can have COB-COL offsets of the same magnitude and direction as our fits. 

Alternatively, a dark pole with an albedo a few times smaller than the global average albedo could similarly produce offsets the size we see---akin to Charon's dark north pole. Recent JWST observations show that C$_2$H$_6$ is present on Quaoar's surface \citep{emery2024tale}.  How would C$_2$H$_6$ behave in Quaoar's surface environment? Sublimation tends to be driven by the maximum temperatures reached. For Quaoar's perihelion distance, a low albedo, low thermal inertia region at the sub-solar point could reach temperatures approaching 63 K. At that temperature, the equilibrium vapor pressure of C$_2$H$_6$ is 6$\times10^{-11}$ bar \citep{fray2009sublimation}, which, according to the Hertz-Knudsen-Langmuir equation \citep{langmuir1913vapor} produces a sublimation rate into vacuum of 0.055 g cm$^{-2}$ year$^{-1}$, or the loss of 1 g cm$^{-2}$ in just 18 years. Velocities of sublimating C$_2$H$_6$ molecules will be governed by the Armand distribution \citep{armand1977classical,schorghofer2022statistical}. Compared to Quaoar's escape velocity of $\sim$540 m s$^{-1}$, these velocities are mostly lower, with only 0.07\% of sublimating C$_2$H$_6$ molecules exceeding escape velocity. The rapid sublimation and low escape fraction suggest that C$_2$H$_6$ should migrate away from equatorial latitudes relatively quickly. Thanks to Quaoar's low obliquity, high latitudes remain much colder, so C$_2$H$_6$ will tend to accumulate there. Over time it should be radiolytically processed into a dark, reddish, high molecular mass residue. Quaoar's current orientation allows us to see a dark south polar region, while hiding much of the north polar region, providing a systematic offset between Quaoar's COL and COB.

\subsection{Possible Systematics}
\label{sec:systematics}
One issue with our detections could be incomplete longitudinal sampling of Quaoar's surface. When looking to our astrometric data (Table \ref{tab:observations}), most of our observations are relatively insensitive to the COB-COL offsets, with the offsets being with 1-2 times the size of typical error bars. In contrast, our last four observations, all from WFC3 in 2024-2025, are much more sensitive to putative COB-COL shifts, with the offsets being $\sim$4-6 times the size of the typical error bars. With observations at just four different longitudes, it may be possible that our sampling of Quaoar's surface may be incomplete. Indeed, taking the times of our last four observations and calculating their rotational phase \citep[with a rotational period of 17.752 hr, ][]{kiss2024visible}, three of the four fall within phases from 0-0.1, just 10\% of all possible longitudes. With this unlucky phasing, our fits may be particularly biased towards a single longitude, providing a skewed measurement of the COB-COL offset. Again comparing to COB-COL offsets for Pluto, sampling at a single phase could produce an offset as much as $\sim$40\% higher/lower than the true mean \citep[see Figure 18 in ][]{buie2010pluto}. Similar issues can come from wavelength dependent COB-COL offsets, which are again important for modeling Pluto's surface. 

Another way to probe the true COB-COL offset is to consider Weywot's eccentricity. As shown in Figures \ref{fig:corner_con} and \ref{fig:corner_uncon}, the COB-COL offset is strongly correlated with fitted eccentricity. Weywot's non-zero eccentricity is interesting when considering the timescale for tidal circularization. Using the circularization timescale from \citet{1966Icar....5..375G}, we find Weywot's circularization timescale is $\lesssim$10 Myr, assuming a range of values for Weywot's density and size. We can then ask, if Weywot is on a circular orbit, what is the required average COB-COL offset to explain our data? Visual inspection of the $e-\Delta x_{col}$ and $e-\Delta y_{col}$ joint posterior distributions in Figures \ref{fig:corner_con} and \ref{fig:corner_uncon} show this occurs near 2-3 mas of total offset. This may be closer to the true COB-COL offset for Quaoar. Offsets of this size are plausible, as they are the same magnitude as those on Pluto ($\sim$5-10\% of the angular diameter). We note that even a circularized Weywot could still retain a small \trackchange{inclination}, as the inclination damping timescales can be orders of magnitude longer than eccentricity damping timescales \citep{1999ssd..book.....M}.


Although we interpret our measured offsets as features on Quaoar's surface, Quaoar's ring system could also contribute to COL variations. A circular homogeneous ring orbiting Quaoar's center-of-mass would provide no net COB-COL, but Quaoar's rings are known to contain denser arcs of material which could provide a net offset. In principle, these offsets may average out over a large number of observations, but as discussed above, our dataset may not be large enough at this point in time. On the other hand, if the outer ring's arc-like structure(s) are confined within Weywot's nearby 6:1 MMR, offsets may be a function of Weywot's orbital phase, providing a systematic bias that may not fully average out. With so little known about Quaoar's ring, we leave a full accounting of their COB-COL effects to future work. 

Likewise, Quaoar's triaxial shape could introduce COB-COL offsets. Although the center of the projected limb of a triaxial body is coincident with the COB, the COL of the visible hemisphere is not necessarily aligned. Depending on the surface reflectance properties, the COL could vary significantly, and may even have systematic offsets. We leave exploration of shape generated COB-COL offsets to future work.

Also of note are possible offsets between Quaoar's center-of-mass (COM) and its COB. Throughout this work, we assumed Quaoar's COB and COM are coincident, but this may not be justified. For example, the Moon has a well-studied COB-COM offset. These offsets, however, will typically be very small. As implied by our low \jt{} measurement, Quaoar's interior likely experienced some degree of melting, allowing hydrostatic relaxation to take place, minimizing COB-COM offsets. Offsets could later develop, possibly after freezing of Quaoar's interior, but would remain small. In any case, our model actually probes COM-COL offsets, so they are fully accounted for in our fits (assuming an averaged constant offset). 

\subsection{How do COB-COL offsets affect other parameters?}
With such uncertainty associated with our measurements of COB-COL offsets, how are our other results impacted? \trackchange{Looking} to Figure \ref{fig:corner_con}, the constrained fit \jt{} measurement is relatively uncorrelated with the offsets, although a slightly smaller $\Delta y_{col}$ implies a slightly larger value of \jt. More broadly, $\Delta x_{col}$ and $\Delta y_{col}$ seem to be primarily correlated with $e$, $\omega$, and $\mathcal{M}$---which is unsurprising given the discussed correlation with $e$---while most other parameters seem to be generally un-/weakly-correlated. Moving to the unconstrained fit, $\Delta x_{col}$ and $\Delta y_{col}$ \trackchange{appear to be mainly correlated with $e$ and orbit angles, although a modest correlation with \jt{} is also present. This correlation with \jt{} is indirect. While \jt{} and COB-COL offsets should not be correlated, both \jt{} and the offsets are correlated with the orbit angles, indirectly making \jt{} and the offsets have a correlation. Indeed, this can be seen when comparing the \jt-$\Delta x/y$ posteriors with the \jt-$\omega$ posteriors, which appear to be mirror images.}

\trackchange{In addition to the fits we present here, we also conducted a series of orbit fits (both constrained and unconstrained) without COB-COL offsets. These fits show remarkable similarity with their COB-COL counterparts, providing nearly identical orbital periods, orbit pole orientation ($i,\Omega$), and measurements of non-Keplerian effects. This strongly suggests the rest of our interpretation is fairly insensitive to constant COB-COL effects.}

\trackchange{If longitudinally variable COB-COL offsets are present, could we be confusing them for non-Keplerian motion? Given Weywot's small eccentricity, the domininant source of non-Keplerian effects will be nodal---rather than apsidal---precession. Nodal precession will primarily cause the orientation of the orbit to subtly change, which would primarily generate motion in the $\Delta y$ direction. With positive/negative offsets on each side of the orbit. In contrast, longitudinal albedo variations will primarily cause $\Delta x$ motion over a Quaoar rotation period. This makes it difficult to confuse nodal precession and varying COB-COL offsets, again suggesting that our results concerning Quaoar's \jt{} are not significantly altered by COB-COL effects.}

\subsection{Moving to a longitudinally variable COB-COL model}
Disentangling Quaoar's COL-COB offsets from Weywot's orbital motion is difficult. The work we present here is just a first-order approximation of the effect, which could be significantly influenced by incomplete sampling of Quaoar's surface. Moving forward, more physically motivated models should be used to better determine these effects. \trackchange{Most importantly, a longitudinally-variable COB-COL offset should be used.}\trackchange{Given the light curve amplitude of Quaoar, our sub-optimal fit quality, and comparison to Pluto and Charon, it is likely that detailed surface maps of Quaoar's are needed to fully explain Weywot's orbital motion.}

\trackchange{Although surface maps are likely more than a decade away, simple implementations of longitudinal variation can be used to jointly understand Weywot's orbit and Quaoar's surface. For example, a simple second order model could use a constant offset and a few Fourier terms to replicate some of the surface albedo variegations. Using techniques similar to those used for Pluto's mutual events, these could plausibly be inverted to provide crude surface maps \citep{buie1992albedo}. Self-consistent modeling of viewing geometry changes from Quaoar and Earth's heliocentric motion should also be included. Alternatively, spherical harmonic expansions could be used to model the longitudinal variations. Various tools have been developed in the exoplanet literature to model starspot's effects on exoplanet transit light curves which could be adapted for this problem \citep[e.g.][]{luger2019starry}.}

\trackchange{Prior to the development of these models, significant work will have to be done to study Quaoar's light curve in more detail. A precise rotation period---which remains elusive for Quaoar---is perhaps the most important measurement to be input into COB-COL models. Without it, phasing all the astrometric observations will be impossible. In addition, a precise light curve can also aid in understanding the COB-COL variations. Unfortunately, Quaoar is currently in a dense star field, which makes photometry challenging.}

\subsection{Other observations to probe COB-COL offsets}
Aside from higher fidelity modeling, observations may also provide detailed constraints. Although no telescopes are currently capable of mapping Quaoar's surface, near- and medium-term future telescopes like 30 m class telescopes or Habitable Worlds Observatory may enable the creation of surface maps, which could correct for Quaoar's COB-COL offset. These are the gold-standard for fully correcting for these effects, but are unfortunately $\sim$decades away.

In the shorter term, measurements of COB-COL variations can be feasibly accomplished by using Weywot as a reference. Although the absolute position of Weywot is somewhat uncertain at any given time (based on our poor fit quality), its orbital motion over a short time is generally well-constrained. This can be seen in our precise measurement of Weywot's orbital period. Repeated observations of Quaoar and Weywot over a single Quaoar rotation period could reveal COB-COL offsets when subtracting out the predictable motion of Weywot. Although requiring $\sim$18 hrs of continuous observing time, this could sensitively detect periodic COB-COL motion. One drawback to this technique is its insensitivity to the average offset. 

Another possible route forward is to use absolute astrometry of Weywot, obviating the need to use Quaoar's COL as an astrometric reference. This technique has been used to study Pluto's small satellites \citep{porter2023orbits}. The HST images we use for relative astrometry typically have at least a few Gaia reference stars in each image which could be used for absolute astrometry. Precise absolute astrometry can also be derived from occultations, which are insensitive to Quaoar's COL. We note that our dataset does include occultation-derived relative astrometry, but these are referenced to the ephemeris of Quaoar which is conditioned upon telescopic observations vulnerable to COB-COL offsets \citep{desmars2015orbit}. Only double occultations---where both Quaoar and Weywot are detected---completely eliminate COB-COL offsets in relative astrometry. Currently, only one double event has been recorded/published \citep[][]{Kretlow2019,fernandez2023weywot}, but occultations of Quaoar and Weywot are relatively common. We defer these tasks to future work and encourage ongoing occultation observations of Quaoar and Weywot.

\begin{figure*}
    \includegraphics[width=\linewidth]{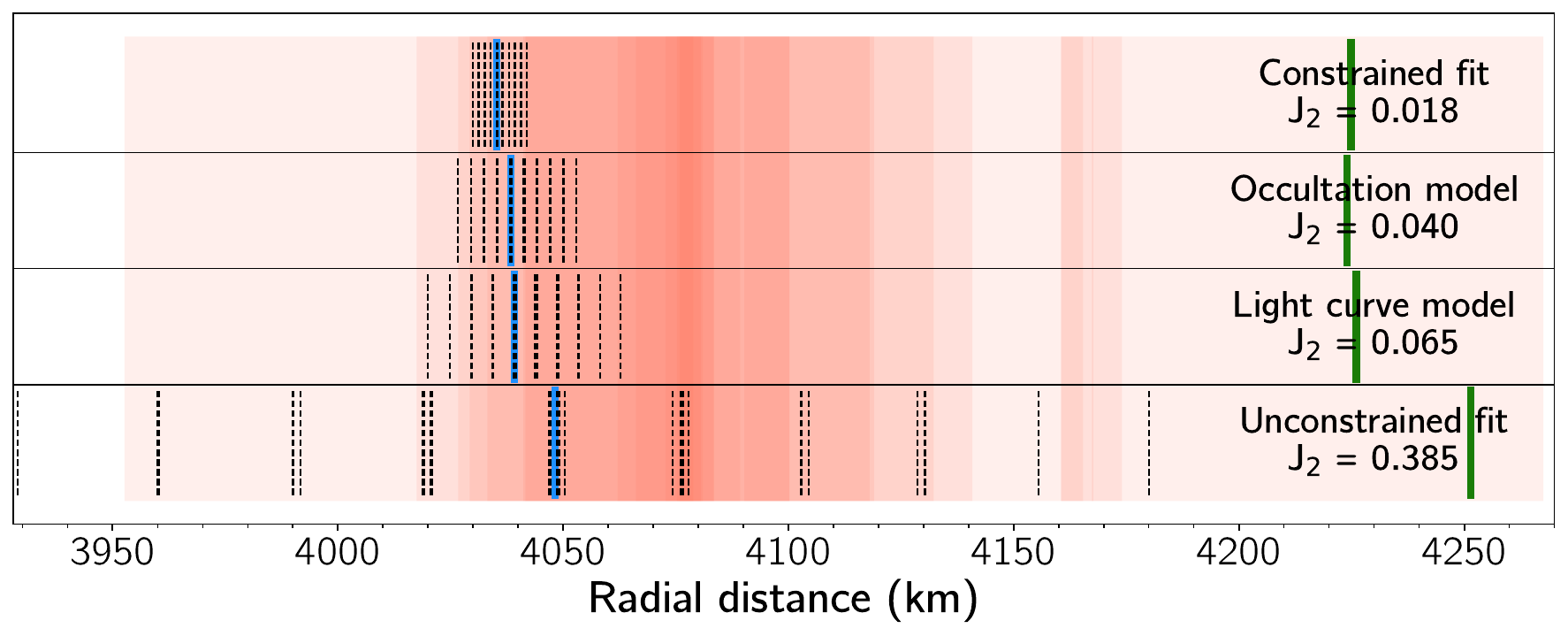}
    \caption{Resonance splitting of Weywot's 6:1 MMR based on the constrained fit, the differentiated occultation model, the differentiated light curve model, and the unconstrained fit. The various 6:1 subresonances are shown as dashed black lines, with the nominal location of the resonance shown in blue. Quaoar's 3:1 SOR is shown with green lines. Ring detections of Q1R \citep[][]{morgado2023dense,pereira2023two,proudfoot2025jwst} are displayed as shaded red regions (with ring radius and ring width), where darker shading is associated with more numerous ring detections. For realistic values of Quaoar's \jt{} (i.e., all except the unconstrained model), Weywot's 6:1 is on the inner edge of the ring, possibly providing a confinement mechanism against inward viscous spreading. }
    \label{fig:subresonances_q1}
\end{figure*}

\section{Ring Dynamics}
\label{sec:rings}

With a better understanding of Weywot's orbit, we are able to better characterize Quaoar's rings. Quaoar's outer ring, Q1R, is near both Weywot's 6:1 MMR, as well as Quaoar's 3:1 SOR. Given Quaoar's \jt, Weywot's 6:1 MMR splits into 28 different fifth-order subresonances \citep{1999ssd..book.....M}. These are made up of the 6 eccentricity-only subresonances explored in past works \citep[e.g.,][]{morgado2023dense, rodriguez2023dynamical}, along with 22 mixed eccentricity and inclination subresonances. In an odd order resonance, there are no inclination-only subresonances at fifth-order--although the tenth-order 12:2 resonance does contain inclination only subresonances. 

Typically, inclination/mixed resonances have been ignored for simplicity, but the substantial inclination of Weywot with respect to the ring system suggest their potential importance. The strength of a k-th order inclination subresonance scales with $\sin{(i/2)}^k$, while eccentricity subresonances scale with $e^k$. As such, Weywot's inclination of $\sim5\degr$ is roughly comparable (in terms of resonance strength) with an eccentricity of $\sim0.04$. This implies that inclination-dominated subresonances may be stronger than eccentricity-dominated/-only subresonances. Inclination resonances are able to excite a variety of interesting phenomena in ring systems. A non-coplanar satellite can induce bending waves and other vertical structure into the rings. Such vertical structure will act to increase the velocity dispersion of ring particles, potentially helping to suppress accretion. 

In Figure \ref{fig:subresonances_q1}, we plot the locations of the 28 subresonances based on our two orbits fits---as well as our two comparison shape models---compared to the ring detections in the literature. Not all of these subresonances will be active, but their radial extent can help better understand the rings. We note that since \jt{} is relatively poorly measured, the subresonance locations will not be precisely measured. Based on the radial locations of subresonances from realistic models (top three rows of Figure \ref{fig:subresonances_q1}), it is clear that the bulk of Q1R is external to Weywot's 6:1 MMR. Based on our constrained fit, the nominal location of the 6:1 is at 4035 km, with subresonances extending $\sim$5-10 km on either size. Given Weywot's low eccentricity, the widths of these resonances are extremely small \citep{rodriguez2023dynamical}. As such, it is unlikely that Q1R is confined within Weywot's 6:1 MMR. Instead, we suggest that the 6:1 plays a role in confining the inner edge of the ring. As the narrow ring spreads radially due to viscous forces (stemming from particle collisions), the 6:1 may act as a barrier to inward spreading. In contrast, the outer edge of Q1R appears to be bounded by Quaoar's 3:1 SOR. Based on our measurement of Quaoar's mass and a rotation period of 17.752 hrs \citep{kiss2024visible}, Quaoar's 3:1 SOR lies at 4225 km. 

In this paradigm, Q1R's arcs remain a mystery. \citet{rodriguez2023dynamical} explored whether Weywot could create arcs within Q1R, finding that particles residing in various subresonances can clump together. The subresonances however, are extremely narrow at Weywot's $e\sim0.01$, with resonance widths expected to be $\lesssim$1 km wide \citep{morgado2023dense}---much larger than the $\sim$5 km wide arcs. Likewise, the arcs appear to be further out from the 6:1 subresonances, limiting their use for explaining Q1R's arcs. One possible alternative is the presence of shepherd moons in/near the ring system, which maintain the ring's eccentricity and create arcs, much like Galatea's influence on Neptune's Adams ring \citep{namouni2002confinement,de2018rings}.

At $\sim$2530 km, Q2R is located near Weywot's 12:1 resonance and Quaoar's 7:5 resonance, at 2543 km and 2541 km respectively. Both of these resonances are high-order, which significantly limits their strength. Detailed simulations should be able to determine at what level these resonances contribute to Q2R's stability/confinement. 

With our updated orbit model for Weywot, high-fidelity simulations of Quaoar's rings are now possible. Future work should focus on understanding the confinement of the rings, modeling both MMRs and SORs simultaneously. Simulations should also reflect the non-coplanarity of the system, which could introduce an interesting variety of phenomena like bending waves and/or warping. Outside of understanding the rings' confinement and location, understanding the origin of Q1R's arcs is particularly important. Tracking of the arc's location could reveal the dynamical source of its confinement, even without detecting individual shepherd moons.

\section{Conclusions}
\label{sec:conclusions}

In this work, we have presented the most comprehensive orbital analysis of the Quaoar–Weywot system to date, leveraging nearly two decades of astrometric data, including new and reanalyzed HST imaging as well as recent stellar occultations. We find the following conclusions:

\begin{compactitem}
    \item In contrast to past analyses, we find that Weywot's orbit is nearly circular with a robust upper limit of $e<0.02$. Weywot is also inclined relative to Quaoar's ring plane, with an inclination of $\sim$5$\degr$. 
    \item Our modeling incorporates Quaoar’s non-spherical gravity field through its \jt{} harmonic. We find relatively small values compared to current shape models of Quaoar, when assuming Quaoar's rings lie in its equatorial plane. \trackchange{We interpret this as confirmation that Quaoar is differentiated and calculate a bulk density of $1751\pm13$ (stat.) kg m$^{-3}$ for Quaoar. When relaxing assumptions about Quaoar and the ring's pole alignment, our model appears to overfit the data, providing unphysically large values of \jt.} 
    \item Our orbit models strongly detect center-of-body to center-of-light (COB-COL) offsets on Quaoar. These offsets are consistent with latitudinal albedo variations on Quaoar’s surface and are necessary to reconcile orbit fits with our dataset of $\sim$milliarcsecond astrometry. Although modeled here as constant offsets, future work should incorporate physically motivated, time-variable models to better capture the effects of surface heterogeneity.
    \item Our measurements of Quaoar's \jt, alongside measurements of Weywot's mean motion and precession rates allows for detailed determinations of Quaoar and Weywot's various resonances. We find that Weywot's 6:1 lies just interior to Quaoar's outer ring, while Quaoar's 3:1 spin-orbit resonance lies just exterior to the ring. We propose that these two resonances provide confinement for the ring, but are unable to explain the dense arcs of material within the ring. Instead, these arcs may be formed by interactions with shepherd satellites in/near the rings.
    \item To facilitate future scientific investigation, we provide an ephemeris of Weywot's position from 2005-2035, which can be used to predict Weywot's position to $\lesssim$10 mas accuracy. 
\end{compactitem}

Our findings show that Weywot serves as a sensitive dynamical probe of Quaoar’s internal structure and shape, enabling constraints on its gravitational field independent of direct shape modeling. These results further inform our understanding of the formation and confinement mechanisms acting on Quaoar’s ring system, and reemphasize the importance of non-Keplerian dynamics in trans-Neptunian satellite systems. Future work---including expanded occultation campaigns, absolute astrometry of Weywot, high-resolution imaging of Quaoar's surface, and direct simulations of rings---will be crucial for disentangling remaining degeneracies and refining models of Quaoar’s interior, surface, and ring system.

\begin{acknowledgments}
We thank the BYU Office of Research Computing for their dedication to providing computing resources without which this work would not have been possible. 

This work is based on observations with the NASA/ESA Hubble Space Telescope obtained from MAST at the Space Telescope Science Institute, which is operated by the Association of Universities for Research in Astronomy, Incorporated, under NASA contract NAS5-26555. Support for Program number 17417 was provided through a grant from the STScI under NASA contract NAS5-26555.

B.P. is supported by the University of Central Florida Preeminent Postdoctoral Program (P$^3$). 

\end{acknowledgments}

\begin{contribution}

B.P. led the overall analysis, writing of the paper, designing/scheduling of observations, and was the principal investigator of the HST program. W.G. analyzed all observations (both new and archival), contributed to interpretation, and provided editing support. D.R. provided access to computing resources, contributed to interpretation, and provided editing support. E.F.V. contributed to occultation interpretation, as well as editing support. 

\end{contribution}

\bibliographystyle{aasjournalv7}
\bibliography{all}

\end{document}